\documentclass[prl, superscriptaddress,widetext, twocolumn]{revtex4}
\usepackage{graphicx}
\usepackage{hyperref}
\def\@biblabel#1{{#1.}} %    %  LV \def\@BIBLABEL#1{$^{#1}\m@th$} %

\def\be{\begin{equation}} \def\ee{\end{equation}}

\newcommand{\ket}[1]{\mbox{$|#1\rangle$}}

\begin{document}

\title{High-fidelity gates in a Josephson qubit}
\author{Erik Lucero}
\affiliation{Department of Physics, University of California at Santa Barbara, Broida Hall, Santa Barbara, CA 93106}
\author{M. Hofheinz}
\affiliation{Department of Physics, University of California at Santa Barbara, Broida Hall, Santa Barbara, CA 93106}
\author{M. Ansmann}
\affiliation{Department of Physics, University of California at Santa Barbara, Broida Hall, Santa Barbara, CA 93106}
\author{Radoslaw C. Bialczak}
\affiliation{Department of Physics, University of California at Santa Barbara, Broida Hall, Santa Barbara, CA 93106}
\author{N. Katz}
\affiliation{Department of Physics, University of California at Santa Barbara, Broida Hall, Santa Barbara, CA 93106}
\affiliation{Department of Physics, Hebrew University, Jerusalem, Israel}
\author{Matthew Neeley}
\affiliation{Department of Physics, University of California at Santa Barbara, Broida Hall, Santa Barbara, CA 93106}
\author{A. D. O'Connell}
\affiliation{Department of Physics, University of California at Santa Barbara, Broida Hall, Santa Barbara, CA 93106}
\author{H. Wang}
\affiliation{Department of Physics, University of California at Santa Barbara, Broida Hall, Santa Barbara, CA 93106}
\author{A. N. Cleland}
\affiliation{Department of Physics, University of California at Santa Barbara, Broida Hall, Santa Barbara, CA 93106}
\author{John M. Martinis}
\email{martinis@physics.ucsb.edu}
\affiliation{Department of Physics, University of California at Santa Barbara, Broida Hall, Santa Barbara, CA 93106}


\date{\today}
\begin{abstract}

We demonstrate new experimental procedures for measuring small
errors in a superconducting quantum bit (qubit). By carefully separating out gate
and measurement errors, we construct a complete error budget and
demonstrate single qubit gate fidelities of 0.98, limited by energy
relaxation. We also introduce a new metrology tool --- a Ramsey
interference error filter --- that can measure the occupation probability of the state $\ket{2}$ 
down to $10^{-4}$, a magnitude near the fault tolerant
threshold. 

\end{abstract}

\maketitle

The immense computational power of a quantum computer comes with a
cost - the fragility of entangled quantum states from coherence
loss. Although decoherence is present in all physical systems, the
effect of the resulting logic errors can be overcome by using error-correcting 
codes, provided that gate errors fall below a fault-tolerance
threshold \cite{Knill2005, Steane1999, Knill1998, Gottesman1998, DiVincenzo1996}. This threshold depends on system architecture and
the specific form of decoherence, but is likely to be $\sim 10^{-4}$
range \cite{Knill2005}. The measurement of gate fidelity in this range is thus a
critical step in implementing fault-tolerant quantum computation. To
date, high fidelity logic gates have only been demonstrated in ion
traps \cite{Leibfried2003,Chiaverini2004}. Solid-state systems such as Josephson qubits \cite{Devoret2004, You2005, Bouchiat1998, Nakamura1999, Vion2002, Chiorescu2003, Makhlin2001, Sillanpaa2007, Majer2007, Plantenberg2007, Niskanen2007, Steffen2006, Martinis2002}, which have the
potential advantage of scalability, have not achieved equivalent fidelities.
Here, we measure the fidelity of a single qubit gate for a
Josephson phase qubit, demonstrating substantial progress towards
this goal. Using the new metrological technique of ``Ramsey
filtering'', we also show how one important error process can be
measured and reduced to the fault-tolerant threshold.

Coherence is typically quantified through the energy decay time $T_1$ and
coherence time $T_2$ (that includes dephasing) obtained from a
Ramsey fringe experiment. The fidelity of a gate operation is then
computed as the ratio of the gate time to coherence time. We note,
however, that such an analysis assumes no loss in fidelity during
a logic gate operation when the quantum state is changed, and thus
it more properly corresponds to the fidelity of a memory operation.
In addition, these coherence times are typically determined by the relative
decay in an experimental signal assumed to be proportional to the
state probability, thus ignoring any fidelity loss that might be
constant in time. A full measurement of gate fidelity, applicable
to the fault-tolerance threshold, should include gate errors that are
determined via probabilities with an absolute calibration.

To illustrate the importance of these issues, we note that many
experimental systems use qubit states $\ket{0}$ and $\ket{1}$, often the ground and first excited states,
chosen from a larger set of basis states \cite{Nielsen2000}. This encoding does not preclude unwanted excitations to other available 
states in the basis. For example, excitations to the next higher energy state $\ket{2}$ are not
necessarily small and correspond to gate
errors that may not be included in standard measurements of $T_1$ and
$T_2$.

In the experiments described here we used a superconducting phase qubit,
where the superconducting phase difference $\delta$ in a Josephson junction 
(with critical current $I_0$) serves as the quantum
variable. When biased close to the critical current, the junction
and its loop inductance $L$ generate a cubic potential where the two
lowest energy eigenstates $\ket{0}$ and $\ket{1}$ have a transition frequency $\omega_{10}/2\pi
\sim 6.75$ GHz (see Fig. 1A). This frequency can be adjusted by $\sim
30$\% using the junction bias current. The circuit layout and operation have been described previously \cite{Steffen2006, Katz2006}.

Single qubit logic operations, corresponding to rotations about the
$x$-, $y$-, and $z$-axes of the Bloch sphere, are generated as
follows: Rotations about the $z$-axis are produced from current
pulses on the qubit bias line that adiabatically change the qubit
frequency, leading to phase accumulation between the states $\ket{0}$ and
$\ket{1}$ \cite{Steffen2006a}. Rotations about any axis in the
$x$-$y$ plane are produced by microwave pulses resonant with the qubit
transition frequency. The phase of the microwave pulses defines the orientation of the
rotation axis in the $x$-$y$ plane, and the pulse duration and
amplitude control the rotation angle.

We perform single shot readout of the phase qubit by applying a fast
($\sim 1\ \textrm{ns}$ rise time) current pulse $I_z$. This fast
pulse lowers the barrier height and increases the tunneling probability of the
$\ket{1}$ state (Fig. 1B). Once tunneled, the state quickly
decays into an external lower energy state that can be easily
distinguished from the untunneled state $\ket{0}$ using an 
on-chip superconducting quantum interference device (SQUID) \cite{Martinis2002}.

\begin{figure}[htb]
\begin{center}
\includegraphics[width=3.5in, trim = 0 3.0in 0 1.5in, clip]{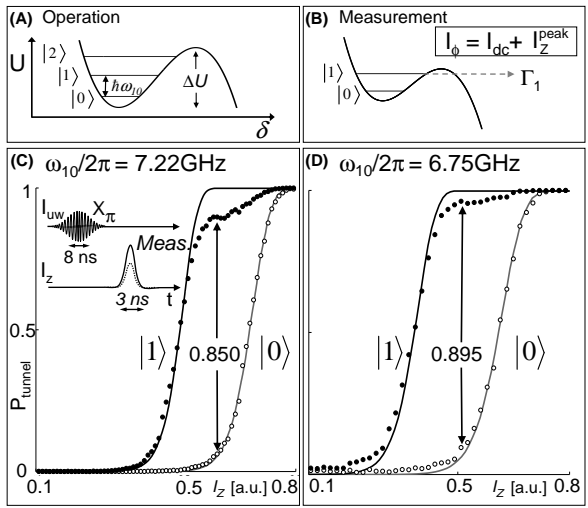}
\end{center}
\caption{Qubit operation and state measurement. ({\bf A}) The potential energy $U$ of a Josephson phase qubit versus junction phase $\delta$. The qubit is formed from the two lowest eigenstates
$\ket{0}$ and $\ket{1}$, with a transition frequency $\omega_{10}/2\pi
\simeq 6.75$ GHz that can be adjusted by varying the dc bias $I_{\phi}=I_{dc}+I_{z}$.
({\bf B}) A measurement pulse lowers the energy barrier $\Delta U$, increasing the $\ket{1}$ state
tunneling probability. ({\bf C}) Tunneling probability versus measurement amplitude $I_z$ for the qubit in the states $\ket{0}$ (open circles) and $\ket{1}$ (filled circles) at qubit frequency $\omega_{10}/2\pi=7.22$ GHz. Fits are shown by the solid lines. ({\bf D}) Data as for {\bf C} but with a larger current bias, $I_{dc}$ giving a smaller qubit transition frequency $\omega_{10}/2\pi=6.75$ GHz.  The visibility between states $\ket{0}$ and $\ket{1}$ is 0.85 and 0.895 for data in {\bf C} and {\bf D}, respectively.  The difference is directly attributed to coupling to a two-level state (TLS) located at $7.05$ GHz, as observed with spectroscopy. The inset illustrates the pulse sequence. The $\ket{1}$  state is prepared by applying a shaped microwave pulse for $\tau = 8$ns, with amplitude chosen to generate a $\pi$ rotation. For the $\ket{0}$ state we apply no microwaves. After state preparation, the current $I_{z}$ is pulsed to measure the qubit state.}
\label{fig:scurve}
\end{figure}

Non-ideal behavior of the qubit can arise from errors either in the
logic gate or in the state measurement. The measurement errors can be accounted for by
thoroughly understanding their physical mechanisms. In Josephson phase qubits, 
measurement fidelities below unity are due
to stray tunneling of the $\ket{0}$ state, the $\ket{1}$ state leaking energy to spurious
two-level states (TLS) \cite{Cooper2004}, and $T_1$ relaxation. To quantitatively
confirm TLS effects as measurement errors, we determined the measurement
fidelity above and below a large TLS splitting at $7.05\ \textrm{GHz}$ (see supplementary material section), as shown in Fig. 1C
and 1D.  For each data set, the tunneling probability of the ground
state $\ket{0}$ is determined versus measurement pulse amplitude
$I_z$. The $X$ pulse is then calibrated for a $\pi$-rotation to give maximum
probability of the $\ket{1}$ state, and the $\ket{1}$ state probability $P_1$ is
determined versus $I_z$. After this calibration, $I_z$ is chosen to give maximum visibility, which is displayed in each figure by an
arrow.

Theoretical predictions for the tunneling probabilities are given by
the solid black and gray lines in Fig. 1C and 1D. The $\ket{0}$ state is misidentified as a $\ket{1}$
state with a probability of $0.034$. This error is consistent with
theory, and corresponds to stray tunneling events during measurement \cite{Cooper2004}. 
At  $\omega_{10}/2\pi =6.75\ \textrm{GHz}$ the $\ket{1}$ state is misidentified as the $\ket{0}$ 
state with a probability of $0.061$, but at a higher qubit frequency, $\omega_{10}/2\pi=7.22\ \textrm{GHz}$ this error increases to $0.106$. The increase in measurement error with qubit frequency is attributed to
 a TLS located between these two frequencies. With a measurement
of the TLS splitting using spectroscopy (see data in supplementary material section), we
predict a $\ket{1}$ state population decrease of $0.045$, a value consistent with our
data. The remaining measurement error is accounted for with an
error budget of $0.010$ for $T_1$ decay, $0.050$ for coupling to
other TLS, and $0.011$ for no tunneling of the $\ket{1}$ state during measurement.

With good agreement between experiment and theory, we can reliably
account for measurement errors in our data. Because the error for the
$\ket{0}$ state --- due solely to stray tunneling --- is simpler and less dependent on systematics, we
choose to perform logic gate experiments that bring the final state
close to $\ket{0}$, thus reducing uncertainties due to state measurement.

The fidelity of a gate is determined by applying two
$\pi$-pulses that produce the transitions $\ket{0} \rightarrow \ket{1}
\rightarrow \ket{0}$, and then measuring the state of the qubit. A
$\pi$-pulse represents the maximum rotation of a single qubit
operation and thus gives a measure of the maximum error for a gate.
Both microwave $\pi$-pulses were designed to have Gaussian envelopes (see supplementary material section), with a duration $8\
\textrm{ns}$ full-width at half maximum (FWHM). The correct
sequential operation of this gate is checked by testing
whether the probability for the final state is independent of the
phase $\Theta$ between the two microwave pulses, as illustrated in
Fig. 2A. In Fig. 2B (2C) the experimental (theoretical) state tomography
data is shown as a function of $\Theta$ and microwave detuning
$\Delta$ from the qubit transition frequency $\omega_{10}/2\pi$. The
experimental data is in excellent correspondence with theoretical
predictions. On resonance ($\Delta = 0$), the phase $\Theta$ has no
effect, as expected, which demonstrates that the two pulses are
calibrated properly as $\pi$-pulses.

\begin{figure}[htb]
\begin{center}
\includegraphics[width = 3.5in, trim = 0 1.5in 0 1.0in, clip]{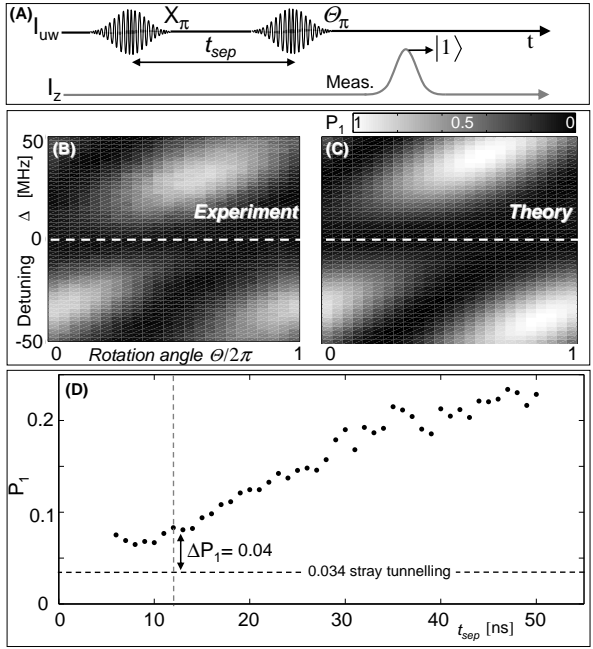}
\end{center}
\caption{Measurement of a high fidelity gate. ({\bf A}) The
pulse sequence consists of two 8 ns Gaussian-shaped $\pi$-pulses, separated in
time by $t_{\rm sep}$, followed by a measure pulse $I_{z}$. The first $\pi$-pulse
defines the rotation axes; by convention this is the x-axis. For the second pulse, which is delayed by $t_{\rm sep}$, we sweep the rotation axis $\Theta$ by changing the phase of the microwaves and detune the microwaves by sideband mixing. This sequence ideally returns the qubit to the $\ket{0}$ state. ({\bf B}) Gray scale plot of measured $\ket{1}$ state probability $P_{1}$ versus detuning $\Delta$ and phase $\Theta$ with $t_{\rm sep} = 12$ ns and ({\bf C}) quantum simulation. On resonance, the phase $\Theta$ does not change $P_{1}$, as expected. ({\bf D}) Plot of $P_{1}$ versus $t_{sep}$. Measurement error of the $\ket{0}$ state is 0.034, as obtained by performing the experiment with no microwaves. The difference between the data and this stray tunneling is 0.04 at $t_{sep} = 12$ ns, corresponding to an error of magnitude 0.02 for each $\pi$-pulse, and a single qubit gate fidelity of 0.98.}
\label{fig:tomo}
\end{figure}
Gate error is directly measured by repeating this experiment with
variable time separation $t_{\rm sep}$ between the two $\pi$-pulses, as
shown in Fig. 2D. The gate error grows with increasing time $t_{\rm sep} > 9\ \textrm{ns}$ because the $\ket{1}$ state decays, and the error has a slope consistent with separate measurements of $T_1$. The error also increases at small times due to the overlap of the two Gaussian microwave pulses. The horizontal dashed line indicates $P_1=0.034$ taken without the application of microwaves; the difference between the data and the dashed line is the gate error. When the pulses are separated by a time $t_{\rm sep} = 12\ \textrm{ns}$, we find an error $\Delta P_1= 0.04$. Since two gate operations are used for this protocol, the fidelity for a single gate operation is $0.98$ \cite{NoteT2}.

\begin{figure}[htb]
\begin{center}
\includegraphics[width = 4.0in, trim = 0 2.0in 1.25in 1.25in, clip]{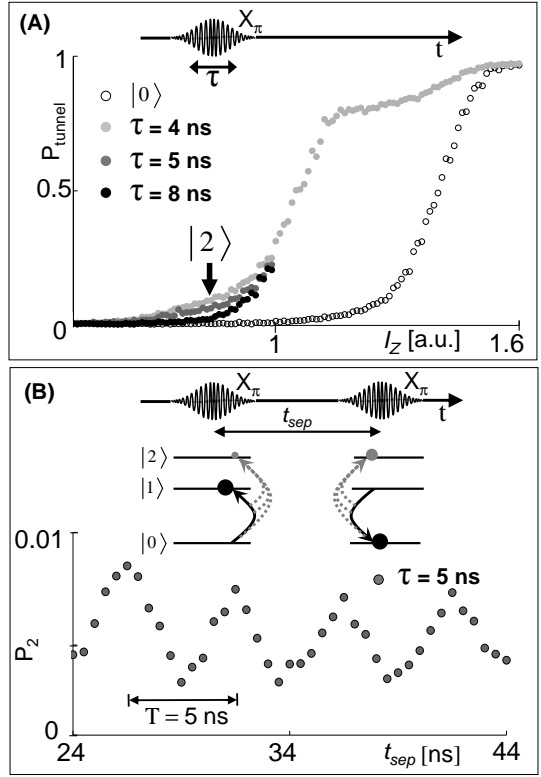}
\end{center}
\caption{Ramsey interference error filter. ({\bf A}) For a singe-pulse sequence,
plot of tunneling probability $P_{tunnel}$ versus $I_{z}$ for $\ket{0}$ and $\ket{1}$ state (gray) for $\tau$ = 4, 5, and 8 ns FWHM Gaussian-shaped $X_{\pi}$-pulses. ({\bf B}) For two-pulse sequence plot of $\ket{2}$ state probability $P_2$ vs. $t_{sep}$ for $\tau$ = 5 ns. The two $X_{\pi}$-pulses are followed by a measure pulse with an amplitude calibrated to tunnel only the
$\ket{2}$ state. During the first $X_{\pi}$-pulse both of the states $\ket{1}$ and $\ket{2}$ are excited. The second $X_{\pi}$-pulse causes the coherent beating of the $\ket{2}$ state. The amplitude of the oscillation is 4 times the error probability, whereas the beat frequency $1/T=1/(5\,\textrm{ns})$ corresponds to the qubit nonlinearity $(\omega_{10} - \omega_{21})/2\pi$. The insets in {\bf A} and {\bf B} illustrate the microwave control pulses. {\bf B} also depicts the three-level system and the unwanted transitions to the $\ket{2}$ state.}
\label{fig:ramsey}
\end{figure}

Initial experiments did not reach this level of performance. We only achieved high fidelity gates by using carefully shaped microwave pulses (see supplementary material section) to minimize excitation of the $\ket{2}$ state \cite{Steffen2003}. There is a tradeoff between using a fast pulse for small $T_1$ errors, or a slow pulse for small Fourier amplitude at the $\ket{1} \rightarrow \ket{2}$ transition frequency, as illustrated in the inset of Fig. 4. The measurement of this error is explicitly shown in Fig. 3A, where $P_{\rm tunnel}$ is plotted versus $I_z$ for a single $\pi$-pulse using $4, 5$, and $8\ \textrm{ns}$ FWHM Gaussian pulses. Excitation to the $\ket{2}$ state produces a shoulder in $P_{\rm tunnel}$ at a value of measurement current $I_z$ below the rise from the $\ket{1}$ state, as indicated by the arrow. This probability is plotted versus Gaussian width $\tau$ in Fig. 4 and shows that this error decreases with increasing pulse width, as expected. Errors become difficult to measure below $\sim 0.01$ because of stray tunneling of the $\ket{1}$ state.

The $\ket{2}$ state error may be measured with much greater sensitivity by recognizing that excitation to the $\ket{2}$ state is a coherent quantum process. Using a two-pulse sequence with variable time delay as illustrated in the inset of Fig. 3B, a Ramsey fringe may be set up between the transitions to the $\ket{2}$ state from the two pulses. We plot in Fig. 3B the $\ket{2}$ state probability $P_2$ versus pulse delay time $t_{\rm sep}$. Since the periodic oscillation is due to coherent interference between the two pulses, the magnitude of this oscillation is four times the probability of exciting the $\ket{2}$ state for a single pulse. More importantly, the ``up-conversion'' of a constant error to an oscillation allows a determination of the amplitude with fewer systematic errors; this error can now be reliably measured down to $10^{-4}$ using this ``Ramsey filter''. The oscillation frequency matches the beat frequency $(\omega_{10}-\omega_{21})/2\pi$ measured via spectroscopy (see supplementary material section), and represents a further check of this measurement technique.

\begin{figure}[htb]
\begin{center}
\includegraphics[width = 4.0in, trim = 0 5.5in 0 1.25in, clip]{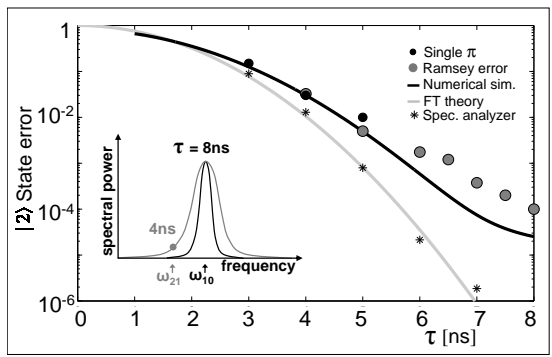}
\end{center}
\caption{Error from $\ket{2}$ state occupation, measured to the fault-tolerant threshold. ({\bf A}) Plot of $\ket{2}$ state error versus Gaussian pulse width for both single $\pi$-pulses (black circles) and Ramsey error (gray circles) data.  The 8 ns FWHM Gaussian produces a $\ket{2}$ probability of $10^{-4}$.  The solid line is the quantum prediction obtained from numerical simulation.  Spectrum analyzer data for the Gaussian shaped pulses (asterisks) are also plotted with the Fourier transform theory curve (solid gray line). The inset illustrates that a $4$ ns pulse produces a significant amount of spectral power at $\omega_{21}/2\pi$.}
\label{fig:fault}
\end{figure}

The $\ket{2}$ state errors determined in this manner are also plotted in Fig. 4. For Gaussian pulses with width 4 and 5 ns, the data from the two methods give a consistent error probability. The error drops exponentially with increasing pulse width, reaching the value $10^{-4}$ at 8 ns. A simple Fourier-transform prediction \cite{Steffen2003} is plotted as a solid gray line, which is computed from the power spectrum of the Gaussian pulse at frequency $\omega_{21}$, normalized to the power at frequency $\omega_{10}$. The asterisks are a measurement of this normalized power taken from the actual control pulses; this simple comparison is an excellent check on the shaping of the microwave pulses as we have found that large spectral leakage gives large qubit error. The solid black line is a prediction of the error obtained from numerical calculations \cite{Steffen2003}, which shows good agreement with the data.

In conclusion, we have demonstrated for single qubits an absolute gate fidelity of 0.98, the highest demonstrated in any solid state system to date. This level of performance was achieved through careful shaping of the microwave control signals. A new metrology tool, Ramsey error filtering, has been introduced, which uses the coherence of an error process for more accurate measurement. We have demonstrated that the probability of the $\ket{2}$ state in our system can be reduced down to $10^{-4}$, a magnitude near the fault-tolerant threshold and that our quantum system remains in the qubit manifold during our single qubit operations. These measurements further demonstrate that superconducting qubits are a leading candidate for a solid-state quantum computer.

%\begin{acknowledgments}
Devices were made at the UCSB and Cornell Nanofabrication Facilities, a part of the NSF-funded National Nanotechnology Infrastructure Network.  This work was supported by ARDA under grant W911NF-04-1-0204 and NSF under grant CCF-0507227.
%\end{acknowledgments}

\bibliographystyle{apsrev}

\begin{thebibliography}{3}

\bibitem{Knill2005} E. Knill, Nature \textbf{434}, 39--44 (2005).

\bibitem{Steane1999} A.~M. Steane, Nature \textbf{399}, 124--126 (1999).

\bibitem{Knill1998} E. Knill,  R. Laflamme, W.~H. ZurekNature, Science \textbf{279}, 342--345 (1998).

\bibitem{Gottesman1998} D. Gottesman, Phys. Rev. A \textbf{57}, 127--137 (1998).

\bibitem{DiVincenzo1996} D.~P. DiVincenzo, P.~W. Shor, Phys. Rev Lett \textbf{77}, 3260--3263 (1996).

\bibitem{Leibfried2003} D. Leibfried \textit{et al.}, Nature \textbf{422}, 412--415 (2003).

\bibitem{Chiaverini2004} J. Chiaverini \textit{et al.}, Nature \textbf{432}, 602--605 (2004).

\bibitem{Devoret2004} M.~H. Devoret, J.~M. Martinis, Quantum Inform. Process. \textbf{3}, 163--203 (2004).

\bibitem{You2005} J.~Q. You, F. Nori, Phys. Today \textbf{58}, 42--47 (2005).

\bibitem{Bouchiat1998} V. Bouchiat, D. Vion, P. Joyez, D. Esteve, M.~H. Devoret, Physica Scripta \textbf{T76}, 165--170 (1998).

\bibitem{Nakamura1999} Y. Nakamura, Y.~A. Pashkin, J.~S. Tsai, Nature \textbf{398}, 786--788 (1999).

\bibitem{Vion2002} D. Vion \textit{et al.}, Science \textbf{296}, 886--889 (2002).

\bibitem{Chiorescu2003} I. Chiorescu, Y. Nakamura,  C.~J.~P.~M. Harmans, J.~E. Mooij, Science \textbf{299}, 1869--1869 (2003).

\bibitem{Makhlin2001} Y. Makhlin, G. Sch\"on and A. Schnirman, Rev. Mod. Phys. \textbf{73}, 357 (2001).

\bibitem{Sillanpaa2007} M. Sillanp\"{a}\"{a}, J.~I. Park and R.~W. Simmonds, Nature \textbf{449}, 438--442 (2007).

\bibitem{Majer2007} J. Majer \textit{et al.}, Nature \textbf{449}, 443--447 (2007).

\bibitem{Plantenberg2007} J.~H. Plantenberg, P.~C. de~Groot, C.~J.~P.~M. Harmans and J.~E. Mooij, Nature \textbf{447}, 836--839 (2007).

\bibitem{Niskanen2007} A.~O. Niskanen, K. Harrabi, F. Yoshihara, Y. Nakamura, S. Lloyd and J.~S. Tsai, Science \textbf{316}, 723--726 (2007).

\bibitem{Steffen2006} M. Steffen \textit{et al.}, Science \textbf{313}, 1423--1425 (2006).

\bibitem{Martinis2002} J.~M. Martinis, S. Nam, J. Aumentado, C. Urbina, Phys. Rev. Lett. \textbf{89}, 117901 (2002).

\bibitem{Nielsen2000} M.~A. Nielsen and I.~L. Chuang, \textit{Quantum Computation and Quantum Information} (Cambridge Univ. Press, 2000).

\bibitem{Katz2006} N. Katz \textit{et al.} Science \textbf{312}, 1498--1500 (2006).

\bibitem{Steffen2006a} M. Steffen \textit{et al.}, Phys. Rev. Lett. \textbf{97}, 050502 (2006).

\bibitem{Cooper2004} K.~B. Cooper \textit{et al.}, Phys. Rev. Lett. \textbf{93}, 180401 (2004).

\bibitem{NoteT2} We note that because of the short duration of these experiments, the coherence was not limited by $T_{2} \simeq 120 $ns.

\bibitem{Steffen2003} M. Steffen, J.~M. Martinis, I.~L. Chuang, Phys. Rev. B. \textbf{68}, 224518 (2003).

\bibitem{Slepian1978} D. Slepian, The Bell system technical journal {\bf 57}, 1371 (1978).

\end{thebibliography}

\section{Supplementary Material}

High-power spectroscopy reveals the transition frequencies between states $\ket{0}$, $\ket{1}$, and $\ket{2}$ and directly measures the nonlinearity of the qubit. The probability of tunneling versus frequency is plotted in the Fig. 5A. The peak at $6.25\ \textrm{GHz}$ corresponds to the qubit $\ket{0} \rightarrow \ket{1}$ transition. The $\ket{1} \rightarrow \ket{2}$ transition is $200\ \textrm{MHz}$ lower in frequency, a value equal to the Ramsey error frequency. For this peak, the $\ket{1}$ state is populated by off-resonant excitation of the $\ket{0} \rightarrow \ket{1}$ transition due to the high power.  A two-photon $\ket{0} \rightarrow \ket{2}$ transition is also observed centered between these two resonances.

\begin{figure}[htb]
\begin{center}
\mbox{\includegraphics*[width=3.5in, trim = 0 2.5in 0 1.25in, clip]{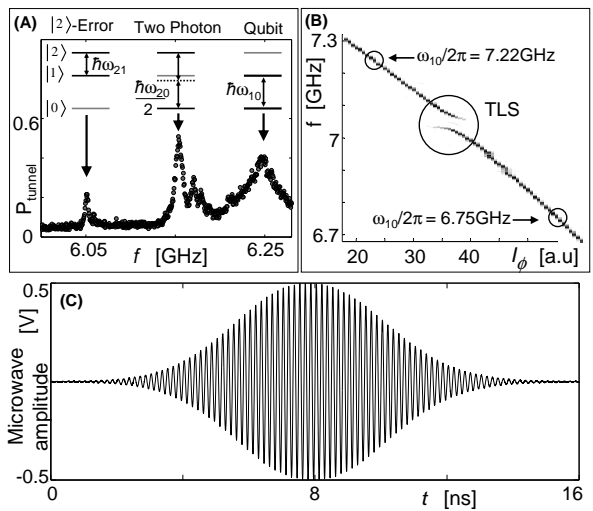}}
\end{center}
\caption{Supplemental data. ({\bf A}) Plot of high power spectroscopy. Inset illustates the transtions. ({\bf B}) Plot of qubit spectroscopy. ({\bf C}) Plot of microwave amplitude versus time.}
\label{Supp}
\end{figure}

The Ramsey error filter data was taken for 4, 5, 6, 6.5, 7, 7.5, and 8 ns FWHM Gaussian pulses. For the longest length pulses, the experiment was repeated $10^{6}$ times.

Qubit spectroscopy is shown in Fig. 5B, where the probability of tunneling is plotted in grayscale for qubit frequency and qubit bias \cite{Cooper2004}. A two-level state (TLS) gives a resonance at $7.05\ \textrm{GHz}$ that couples to the qubit with splitting size $50\ \textrm{MHz}$  The qubit was operated above ($7.22\ \textrm{GHz}$) and below ($6.75\ \textrm{GHz}$) the TLS resonance.

Shown in  Fig. 5C is an example of a Gaussian-shaped microwave pulse taken with a high-speed sampling oscilloscope. These pulses have nearly ideal spectral quality, and are significantly improved compared to those used in previous experiments \cite{Katz2006}.  They are created with a continuous microwave source controlled by an IQ mixer fed by dual 1 GHz digital to analog converters (DAC). The microwave source drives in saturation the local oscillator input of the mixer at frequency $f_0$. The DAC channels are generated in a custom board using AD9736 chips that have 14 bit resolution.  They drive the I and Q ports through $250\,{\rm MHz}$ ($-3\,{\rm dB}$ frequency) dissipative Gaussian lowpass filters and low distortion differential amplifiers.  The microwave output of the mixer is filtered by a 7 pole Chebyshev lowpass filter at $8.5\,{\rm GHz}$ to suppress harmonics of $f_0$.  The large bandwidth of the control signal allows for sideband mixing.  By applying sine and cosine waves at $f_{\rm sb}$ to the I and Q ports, the mixer generates an output signal at frequency $f_0 + f_{\rm sb}$.  Sideband mixing allows for very high on/off ratios of qubit control since the 
(small) carrier leakage at $f_0$ is off resonance with the qubit.  The digital control allows imperfections of the DAC chain and the IQ mixer to be corrected by first measuring its response function and then correcting it with deconvolution.  The relative amplitudes and phases of the I and Q mixer channels are calibrated by minimizing the power at the opposite sideband $f_0 - f_{\rm sb}$.  This is done at enough sideband frequencies so that all Fourier component of an arbitrary digital input signal can be corrected.  In total, we obtain accurate pulse shapes with greater than $60\,{\rm dB}$ suppression of spurious frequencies and harmonics.

For the gate fidelity measurements, the shape of the control pulses were Slepian \cite{Slepian1978}. These pulses have similar envelopes to Gaussians, but have tails that are strictly set to zero.

\end{document}